\documentclass{eccm-ecfd}

\usepackage{graphicx}
\usepackage{amsmath}
\usepackage{amsfonts}
\usepackage{amssymb}
\usepackage{subfigure}
\usepackage{color}

\title{MIXED MODELING FOR LARGE-EDDY SIMULATION: THE MINIMUM-DISSIPATION-BARDINA-MODEL}

\author{LARISSA B. STREHER, MAURITS H. SILVIS AND ROEL VERSTAPPEN}

\heading{Larissa B. Streher, Maurits H. Silvis and Roel Verstappen}

\address{Johann Bernoulli Institute for Mathematics and Computer Science,  University of Groningen\\
P.O. Box 407, 9700 AK Groningen, The Netherlands\\
E-mail: \{l.b.streher, m.h.silvis, r.w.c.p.verstappen\}@rug.nl}


\abstract{The Navier-Stokes equations describe the motion of viscous fluids. In order to predict turbulent flows with reasonable computational time and accuracy, these equations are spatially filtered according to the large-eddy simulation (LES) approach. The current work applies a volume filtering procedure according to Schumann \cite{Schumann_1975}. To demonstrate the procedure the Schumann filter is first applied to a convection-diffusion equation. The Schumann filter results in volume-averaged equations, which are not closed. To close these equations a model is introduced, which represents the interaction between the resolved scales and the small subgrid scales. Here, the anisotropic minimum-dissipation model of Rozema et al.\ \cite{Rozema_2015} is considered. The interpolation scheme necessary to evaluate the convective flux at the cell faces can be viewed as a second filter. Thus, the convection term of the filtered convection-diffusion equation can be interpreted as a double-filtered term. This term is approximated by the scale similarity model of Bardina et al.\ \cite{Bardina_1983}. Thus, a mixed minimum-dissipation-Bardina model is obtained. Secondly, the mathematical methodology is extended to the Navier-Stokes equations. Here, the pressure term is analyzed separately and added to the convection-diffusion equation as a sink term. Finally, spatially filtered Navier-Stokes equations that depend on both the anisotropic minimum-dissipation (AMD) model proposed by Rozema et al.\ \cite{Rozema_2015} and the scale similarity model of Bardina et al.\ \cite{Bardina_1983} are obtained. Hence, a mathematically consistent method of mixing the AMD model and the Bardina model is achieved.}

\begin{document}

\thispagestyle{empty}

\section{INTRODUCTION}

The continuous downscaling of integrated circuits has led to the ongoing increase of the available computational power. Although this growth enables the direct computation of increasingly complicated flows, most practical turbulent flows still cannot be directly computed from the Navier-Stokes equations. Therefore, simulation approaches that reduce the number of degrees of freedom are in constant development.

The current work is based on the large-eddy simulation LES approach. This method aims for the best compromise between required computational time and achieved accuracy through a spatial filtering of the Navier-Stokes equations. The turbulence spectrum is, then, divided into large and small scales of motion. The former are directly computed, while the latter are modeled. 

The filtering process is usually carried out by the application of a convolution filter. This work applies the volume balance procedure proposed by Schumann \cite{Schumann_1975}. In this approach, the governing equations, i$.$e$.$, conservation of mass and momentum, are implicitly filtered according to a Schumann box filter, which is equivalent to the process of averaging these equations over the grid volumes. Like the conventional filtering approach, the application of the Schumann box filter leads to a new term in the momentum equation: the subgrid-scale stress tensor. This term accounts for the effect of the subgrid modes on the resolved part of the flow, that is, on the large eddies. However, the subgrid-scale stress tensor is not closed, since it is not solely expressed in terms of resolved quantities. Therefore, a closure model is needed before actual large-eddy simulations can be performed. This closure is achieved by modeling the subgrid-scale stress tensor according to functional, structural or mixed models \cite{Sagaut_2006}.

Functional models aim at correctly predicting the kinetic energy transfer between the resolved and subgrid eddies. Although these models can produce the correct average energy removal from the large eddies, they poorly represent the effects of the small eddies on the large ones on a local basis. Moreover, they do not describe accurately the structure of the subgrid-scale stress tensor, leading to low correlations between exact and computed subgrid-scale stress tensors \cite{Sagaut_2006} (see Clark et al. \cite{Clark_1979} and Bardina et al. \cite{Bardina_1983}).

Eddy viscosity models are the most applied functional models and are usually easily implemented. A dissipative term, which has a similar effect as the subgrid-scale stress tensor, is introduced. This term is based on the Boussinesq approximation \cite{Boussinesq_1877}, which establishes the concept of an eddy viscosity. The Smagorinsky \cite{Smagorinsky_1963} and the anisotropic minimum-dissipation \cite{Rozema_2015} models are examples of eddy viscosity models.

Structural models, on the other hand, aim at a mathematical reconstruction of the subgrid stress tensor. They are based on formal series expansions of the unknown terms, transport equations, approximative deconvolution procedures or scale similarity arguments in order to achieve a high-fidelity representation of the subgrid-scale stress tensor, also with regard to the eigenvectors of the exact tensor \cite{Sagaut_2006}. Therefore, these models are characterized by higher correlations between the exact and the approximated unresolved stress tensors when compared to functional models (see Bardina et al.\ \cite{Bardina_1983}).

Scale similarity models are structural models that are frequently applied to approximate the subgrid-scale stress tensor. These models are based on the similarity between neighboring energy bands in the inertial subrange \cite{Bensow_2007}, assuming that the turbulent kinetic energy of the unresolved scales can be approximated by the turbulent kinetic energy contained in the smallest resolved eddies. Although scale similarity models such as the Bardina model \cite{Bardina_1983} are characterized by high local correlations between the modeled and the exact subgrid-scale stress tensor, these models do not dissipate enough energy. Hence, they are often supplemented with an eddy viscosity model, forming a mixed model.  

Mixed models are based on a linear combination of functional and structural models. This type of model generally arises from an \textit{ad hoc} combination of an eddy viscosity model with a scale similarity model, in order to counteract specific model limitations. In particular, mixed models are aimed at obtaining good results on both the energy and structural levels \cite{Sagaut_2006}. For instance, Bardina et al.\ \cite{Bardina_1983} proposed the linear combination of their scale similarity model with the Smagorinsky model \cite{Smagorinsky_1963}, achieving a good representation of the local subgrid-scale stress tensor, as well as the mean energy balance for homogeneous isotropic turbulence and homogeneous turbulence in the presence of mean shear. Moreover, Vreman et al.\ \cite{Vreman_1997} simulated a weakly compressible temporal mixing layer with solely structural and functional models, as well as with mixed models. Improved results were achieved for the combination of the Bardina model with the dynamic Smagorinsky model (see Germano \cite{Germano_1992}).

Existing mixed models mostly resulted from an \textit{ad hoc} process of combining eddy viscosity and scale similarity models. This process, however, is not consistent with the derivation of these models \cite{Bensow_2007}. Hence, this work aims to provide the basis for a mathematically consistent combination of eddy viscosity and scale similarity models. In particular, the combination of the scale similarity model of Bardina et al.\ \cite{Bardina_1983} with the anisotropic minimum-dissipation model (AMD) of Rozema et al.\ \cite{Rozema_2015} is carried out. 

The scale similarity model of Bardina et al.\ \cite{Bardina_1983} is based on an extrapolation from the smallest resolved scales, achieved by a double spatial filtering, and is able to capture anisotropic and out-of-equilibrium effects. The application of this model leads to high correlations between the exact and the computed subgrid-scale stress tensor. For instance, correlations as high as $0.8$ on the tensor level were achieved for both homogeneous isotropic turbulence at a subgrid Reynolds number of $Re_{SGS}=180$ and homogeneous turbulence in the presence of mean shear at $Re_{SGS}=204$ (see Bardina et al.\ \cite{Bardina_1983}). Despite of the characteristic high local correlations, the Bardina model does not ensure a positive net rate of energy transfer to the small scales. Hence, this scale similarity model is often combined with an eddy viscosity model. 

The eddy viscosity model that is applied in conjunction with the Bardina model \cite{Bardina_1983} in the current work, i$.$e$.$, the anisotropic minimum-dissipation model \cite{Rozema_2015}, is based on a decoupling of the small scales from the large ones. In order to achieve this scale separation, an eddy dissipation is introduced through the application of a Poincar\'e inequality. This eddy dissipation counterbalances the production of small eddies, removing the energy of the subgrid scales of the solution \cite{Rozema_2015}. The described procedure ensures that the unresolved scales do not influence the solution (see Verstappen \cite{Verstappen_2016}) and consequently allows the formulation of a closure model based only on the resolved eddies. 

Although the eddy dissipation introduced by the anisotropic minimum-dissipation mo-\break del has a similar effect as the subgrid-scale stress tensor, the model does not necessarily have the same structure as the subgrid stresses. For instance, considering other eddy visco-\break sity models such as the Smagorinsky model, correlations between the exact and the computed subgrid-scale stress tensor as low as $0.26$ and $0.03$ were achieved. The former is related to homogeneous isotropic turbulence at low Taylor micro-scale Reynolds numbers ($Re_\lambda<40$) (see Clark et al.\ \cite{Clark_1979}), while the latter is related to homogeneous turbulence in the presence of mean shear at a subgrid-scale Reynolds number of $Re_{SGS}=204$ (see Bardi-\break na et al.\ \cite{Bardina_1983}). Moreover, the anisotropic minimum-dissipation model cannot account for the\break effect of backscatter other than through the introduction of a negative eddy viscosity. Ho-\break wever, that may lead to numerical instabilities in simulations (see Bensow and Fureby \cite{Bensow_2007}).  

Based on the properties of the scale similarity model of Bardina et al.\ \cite{Bardina_1983} and the anisotropic minimum-dissipation model of Rozema et al.\ \cite{Rozema_2015}, it is clear that they are of complimentary nature. Therefore, a mathematically consistent combination of both models is promising. The main goal of this work is to obtain in a natural way the minimum-dissipation-Bardina model.

The current paper is organized as follows: Section \ref{sec:Sem_nome_por_enquanto} provides a thorough description of the applied mathematical methodology in the context of the convection-diffusion equation. Afterwards, in Section \ref{sec: Mixed_modeling}, this mathematical methodology is extended to the Navier-Stokes equations. This results in the incompressible spatially filtered Navier-Stokes equations that includes both the anisotropic minimum-dissipation model \cite{Rozema_2015} and the Bardina\break model \cite{Bardina_1983}. Hence, a mathematically consistent mixed model, i$.$e$.$, the minimum-dissipation-Bardina model, is obtained for large-eddy simulations. Finally, in Section \ref{sec: Conclusions}, the current work is summarized.

\section{MIXED MODELING: MATHEMATICAL METHODOLOGY}
\label{sec:Sem_nome_por_enquanto}

In the current work, a mathematical methodology to obtain a natural combination of the scale similarity model of Bardina et al.\ \cite{Bardina_1983} and the anisotropic minimum-dissipation model of Rozema et al.\ \cite{Rozema_2015} is proposed. To demonstrate this methodology, a two-dimensio-\break nal convection-diffusion equation is analyzed in this section. Section \ref{sec: Mixed_modeling} is devoted to the extension of the proposed methodology to the full three-dimensional Navier-Stokes equations.

The convection-diffusion equation describes the transport phenomena due to convection and diffusion processes:
\vskip-.6cm
\begin{eqnarray}
\label{eq:convection_diffusion}
\frac{\partial f_i}{\partial t}+\frac{\partial f_i\,u_j}{\partial x_j}=D\,\frac{\partial^2\, f_i}{\partial x_j\,\partial x_j}\,.
\end{eqnarray}
The first and second terms on the left-hand side come from the material derivative of the variable $f_i$ and represent the time variation and the convective transport, respectively. The term on the right-hand side represents the diffusion. The density of any physical variable is represented by $f_i$. Although the density $f_i$ is currently represented as a vector, it can also be a scalar. The velocity field with which the physical variable is convected and the diffusion coefficient are represented by $u_j$ and $D$, respectively. Einstein's summation convention is implied for repeated indices.

The convection-diffusion equation (Eq.\ (\ref{eq:convection_diffusion})) is spatially filtered with the help of the Schumann \cite{Schumann_1975} filter, which is defined by
\vskip-.6cm
\begin{eqnarray}
\label{eq:Schumann_box_filter_1}
 ^{^{^V}}\!\overline{f}_i=\frac{1}{|V|}\int_{V}f_i\,\,dV\,,
\end{eqnarray}
where V denotes the volume of the filter box. The obtained volume-averaged convective and diffusive terms are rewritten by applying Gauss' divergence theorem. This procedure leads to the appearance of surface-averaged terms, which are defined by
\vskip-.6cm
\begin{eqnarray}
\label{eq:surface_average}
^{^{^S}}\overline{f}_i=\frac{1}{|S|}\int_{S}f_i\,\,dS\,,
\end{eqnarray}
where S denotes a surface (e$.$g$.$ the surface of $V$). Thus, the volume-averaged convection-diffusion equation becomes
\vskip-.6cm
\begin{eqnarray}
\label{eq:volume_averaged_convection_diffusion}
\frac{|V|}{|S|}\,\frac{\partial\, ^{^{^V}}\!\overline{f}_i}{\partial t}+\,^{^{^S}}\overline{f_i\,u}_j=^{^{^{^{^{^S}}}}}\!\!\overline{D\,\frac{\partial\,f_i}{\partial x_j}}\,.
\end{eqnarray}
This equation is, however, not closed due to the nonlinearity of the convective term, i$.$e$.$, the second term on the left-hand side of Eq.\ (\ref{eq:volume_averaged_convection_diffusion}). Therefore, this term is decomposed according to
\vskip-.6cm
\begin{eqnarray}
\label{eq:surface_averaged_convection_term}
^{^{^S}}\overline{f_i\,u}_j=^{^{^V}}\!\!\!\overline{f}_i\,^{^{S}}\overline{u}_j+\tau_{ij}^{\alpha}\,,
\end{eqnarray}
where $\tau_{ij}^\alpha$ represents the subgrid-scale stress tensor. In this work, this tensor is determined according to an eddy viscosity approach:
\vskip-.6cm
\begin{eqnarray}
\label{eq:subgrid_scale_stress_tensor}
\tau_{ij}^{\alpha}\approx-^{^{^{^{^{^S}}}}}\!\overline{D_e\,\frac{\partial f_i}{\partial x_j}}\,.
\end{eqnarray}
Thus, the nett effect of the subgrid stress in the filtered convection-diffusion equation is an increase of the diffusion coefficient; The total diffusion coefficient becomes $D+D_e$, where $D_e$ is the diffusion coefficient related to the turbulence.

\begin{figure}[h]
	\centering
	\def\svgwidth{6cm}
	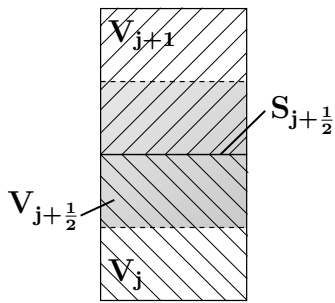
	\caption{Shifted volume in relation to the $j\,$-direction. $S_{j+1/2}$ denotes the surface that is included in the volume $V_{j+1/2}$.}
	\label{fig:shifted_volume_j_direction}
\end{figure}

The left-hand side of Eq.\ (\ref{eq:surface_averaged_convection_term}) contains a surface integral, whereas the right-hand side contains a volume integral. To approximate both, shifted volumes are to be introduced. These volumes have the same size and form as the original volumes but are shifted so that they are centered around a surface. Figure \ref{fig:shifted_volume_j_direction} illustrates a volume shifted in the j-direction with regard to a two-dimensional cell.

Obviously the fluxes through all cell surfaces must be determined. Here, the focus is solely on the computation of the convective term at the surface $S_{j+1/2}$ for the sake of brevity. This surface is the intersection of the $V_j$ and $V_{j+1}$ volumes, i$.$e$.$, $V_j\cap V_{j+1}$ (see Fig.\ \ref{fig:shifted_volume_j_direction}).

Firstly, the $^{^{^V}}\!\overline{f}_i$ term in Eq.\ (\ref{eq:surface_averaged_convection_term}) is to be evaluated at the surface $S_{j+1/2}$. For that, the volume average regarding the shifted volume is considered (see Fig.\ \ref{fig:shifted_volume_j_direction}). It is approximated according to
\vskip-.6cm
\begin{eqnarray}
\label{eq:shifted_volume_averaged_f}
^{^{^{V_{j+1/2}}}}\overline{f}_i\,=\,^{^{^{V_{j}\,\cup\, V_{j+1}}}}\overline{f}_i+r_i{|_{_{_{_{S_{j+1/2}}}}}}\,,
\end{eqnarray} 
where $^{^{^{V_{j}\,\cup\, V_{j+1}}}}\overline{f}_i$ represents the volume average of $f_i$ over the volume consisting of the union of the $V_j$ and $V_{j+1}$ cells, i$.$e$.$, ${{{V_{j}\cup V_{j+1}}}}$, and $r_i$ describes the residual at the considered surface.
  
The first term on the right-hand side of Eq.\ (\ref{eq:shifted_volume_averaged_f}) is computed by interpolating the known volume averages of the physical variable $f_i$:
\vskip-.6cm
\begin{eqnarray}
\label{eq:shifted_volume_averaged_union}
^{^{^{V_{j}\,\cup\, V_{j+1}}}}\overline{f}_i=\frac{1}{2}\left(^{^{^{V_{j}}}}\overline{f}_i+^{^{^{V_{j+1}}}}\overline{f}_i\right).
\end{eqnarray}

Equation (\ref{eq:shifted_volume_averaged_union}) shows that the interpolation of $^{^{^{V_{j}}}}\overline{f}_i$ and $^{^{^{V_{j+1}}}}\overline{f}_i$ can be seen as a filter over the volume ${{{V_{j}\,\cup\, V_{j+1}}}}$ (see Fig.\ \ref{fig:shifted_volume_j_direction}). Hence, $^{^{{V_{j}\,\cup\, V_{j+1}}}}\overline{f}_i$ is considered a double-filtered variable. The first filter level is, then, characterized by the same filter width as the Schumann filter, i$.$e$.$, $V_j$ or $V_{j+1}$. The second filter level is characterized by a double filter width in regard to the Schumann filter, i$.$e$.$, a volume filter over $V_j\cup V_{j+1}$. The mathematical methodology, then, naturally introduces a relation between a single-filtered variable, i$.$e$.$, $^{^{{V_{j+1/2}}}}\overline{f}_i$, and a double-filtered variable, i$.$e$.$, $^{^{{V_{j}\,\cup\, V_{j+1}}}}\overline{f}_i$. Therefore, a scale similarity model is the natural choice to approximate the residual $r_i$. This type of model is based on the scale similarity hypothesis, which states that the effect of the unresolved scales on the resolved ones can be approximated through the similarity of the smallest resolved scales and the biggest unresolved scales. This leads to
\vskip-.6cm
\begin{eqnarray}
\label{eq:scale_similarity_hypothesis}
f_i'\approx \overline{f'}_{\!\!i} = \overline{f}_i - \widetilde{\overline{f}}_i \,,
\end{eqnarray}
where $f_i'$ is defined by $f_i=\overline{f_i}+f_i'$. The first and second filter levels are characterized respectively by the filter widths $\overline{\Delta}$ and $\widetilde{\Delta}$, where $\widetilde{\Delta}>\overline{\Delta}$.
It may be remarked that Eq.\ (\ref{eq:scale_similarity_hypothesis})\break applies to a volume filter, as well as to a surface filter. Thus, the residual $r_i$ in Eq.\ (\ref{eq:shifted_volume_averaged_f}) can be defined as
\vskip-.6cm
\begin{eqnarray}
\label{eq:residual_r}
r_i{|_{_{_{_{S_{j+1/2}}}}}}=^{^{^{V_{j+1/2}}}}\!\!\overline{f'}_{\!\!i}\,. 
\end{eqnarray}

\begin{figure}[h]
	\centering
	\def\svgwidth{6cm}
	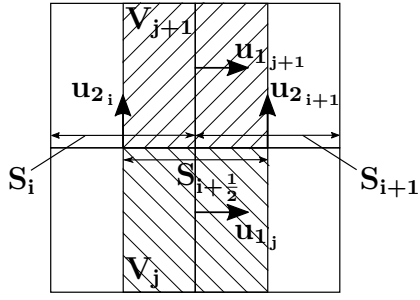
	\caption{Staggered grid: Surfaces and velocities.}
	\label{fig:velocity_average}
\end{figure}

Secondly, the velocity component of the convective term (see Eq.\ (\ref{eq:surface_averaged_convection_term})), i$.$e$.$, $^{^{S}}\overline{u}_j$, is defined at the surface $S_{j+1/2}$. For staggered grids $^{^{S_{j+1/2}}}\overline{u}_j$ must be approximated by interpolation:
\vskip-.6cm
\begin{eqnarray}
\label{eq:surface_average_sj+1/2}
^{^{{S_{j+1/2}}}}\overline{u}_j\,=\,^{^{{S_{i}\,\cup\, S_{i+1}}}}\overline{u}_j+q_j{|_{_{_{_{S_{j+1/2}}}}}}\,,
\end{eqnarray}
where $\,^{^{{S_{i}\,\cup\, S_{i+1}}}}\overline{u}_j$ represents the surface average of $u_j$ over the surface consisting of the union of the $S_i$ and $S_{i+1}$ surfaces (see Fig.\ \ref{fig:velocity_average}). The residual of of this approximation ($q_j$) is to be defined at the $S_{j+1/2}$ surface. For the sake of simplicity, double indices are abolished and only the essential index is shown. For instance, the $j$-index is abolished for the variables located at $j+1/2$. Thus the interpolation of $\,^{^{{S_{i}\,\cup\, S_{i+1}}}}\overline{u}_2$ can be written as:
\vskip-.6cm
\begin{eqnarray}
\label{eq:surface_average_union}
^{^{{S_{i}\,\cup\, S_{i+1}}}}\overline{u}_2=\frac{1}{2}\left(^{^{{S_{i}}}}\overline{u}_2+^{^{{S_{i+1}}}}\!\overline{u}_2\right)\,.
\end{eqnarray}

As for the volume averages of $f_i$, the applied interpolation is interpreted as a filtering process characterized by a filter width of ${{{S_{i}\cup S_{i+1}}}}$. Hence, $^{{{S_{i}\,\cup\, S_{i+1}}}}\overline{u}_2$ is also considered a double-filtered variable, where the first and second filter levels are characterized by filter widths of $S_{i}$ or $S_{i+1}$ and ${{{S_{i}\cup S_{i+1}}}}$, respectively. Again, a natural relation between a single-filtered variable, i$.$e$.$, $^{^{{S_{j+1/2}}}}\overline{u}_2$, and a double-filtered variable, i$.$e$.$, $^{^{{S_{i}\,\cup\, S_{i+1}}}}\overline{u}_2$, is achieved. Therefore, a scale similarity model (see Eq.\ \ref{eq:scale_similarity_hypothesis}) is also a natural choice to model the residual $q_j$. The model is, then, defined according to 
\vskip-.6cm
\begin{eqnarray}
\label{eq:residual_q}
q_j{|_{_{_{_{S_{j+1/2}}}}}}=^{^{^{S_{j+1/2}}}}\!\!\overline{u'}_{\!j}\,.
\end{eqnarray} 

Equations (\ref{eq:shifted_volume_averaged_f}) and (\ref{eq:surface_average_sj+1/2}) are, then, introduced in Eq.\ (\ref{eq:surface_averaged_convection_term}) to approximate the convective flux trough the surface $S_{j+1/2}$:
\vskip-.6cm
\begin{eqnarray}
\label{eq:convective_term_bardina}
^{^{^{S_{j+1/2}}}}\overline{f_i\,u}_j=\,^{^{^{V_{j}\,\cup\, V_{j+1}}}}\overline{f}_i\,^{^{{S_{i}\,\cup\, S_{i+1}}}}\overline{u}_j+\tau_{ij}^{\alpha}{|_{_{_{_{S_{j+1/2}}}}}}+\tau_{ij}^{\beta}{|_{_{_{_{S_{j+1/2}}}}}}\,,
\end{eqnarray}
with
\vskip-.6cm
\begin{eqnarray}
\label{eq:subgrid_tensor_beta}
\tau_{ij}^{\beta}{|_{_{_{_{S_{j+1/2}}}}}}=^{^{^{V_{j}\,\cup\, V_{j+1}}}}\overline{f}_i\,q_j{|_{_{_{_{S_{j+1/2}}}}}}+^{^{{S_{i}\,\cup\, S_{i+1}}}}\overline{u}_j\,r_i{|_{_{_{_{S_{j+1/2}}}}}}+\left(r_i\,q_j\right){|_{_{_{_{S_{j+1/2}}}}}}\,.
\end{eqnarray}

The scale similarity approximations for the residuals $r_i$ and $q_j$ at the surface $S_{j+1/2}$ (see Eqs.\ (\ref{eq:residual_r}) and (\ref{eq:residual_q})) are introduced in Eq.\ (\ref{eq:subgrid_tensor_beta}). This results in:
\vskip-.6cm
\begin{eqnarray}
\label{eq:subgrid_tensor_beta_Bardina}
\tau_{ij}^{\beta}{|_{_{_{_{S_{j+1/2}}}}}}=^{^{^{V_{j+1/2}}}}\overline{f}_i\,^{^{{S_{j+1/2}}}}\overline{u}_j-^{^{^{V_{j}\,\cup\, V_{j+1}}}}\overline{f}_i\,^{^{{S_{i}\,\cup\, S_{i+1}}}}\overline{u}_j\,.
\end{eqnarray}

Here, we only considered the surface $S_{j+1/2}$. The convective term is treated in a similar fashion for all surfaces. Once again, non-relevant indices are supressed as much as possible. Moreover, the double-filtered variables such as $^{^{^{V_{j}\,\cup\, V_{j+1}}}}\overline{f}_i$ and $^{^{{S_{i}\,\cup\, S_{i+1}}}}\overline{u}_j$ are simply denoted by $^{^{^V}}\widetilde{\overline{f}}_i$ and $^{^{S}}\widetilde{\overline{u}}_j$, respectively. And the single-filtered variables such as $^{^{^{V_{j+1/2}}}}\overline{f}_i$ and $^{^{{S_{j+1/2}}}}\overline{u}_j$ will be denoted by $^{^{^V}}\!\overline{f_i}$ and $^{^{S}}\overline{u_j}$, respectively. Hence, the resulting spatial-filtered convection-diffusion equation is
\vskip-.6cm
\begin{eqnarray}
\label{eq:volume_averaged_convection_diffusion_resolved_}
\frac{\partial\, ^{^{^V}}\!\overline{f}_i}{\partial t}+\delta_j\left(\,^{^{^{V}}}\!\widetilde{\overline{f}}_i\,^{^{{S}}}\widetilde{\overline{u}}_j\right)=\delta_j\left(\,^{^{^{^{^{^{S}}}}}}\!\overline{D\,\frac{\partial\,f_i}{\partial x_j}}\,\right)-\delta_j\left(\tau_{ij}^{\alpha}+\tau_{ij}^{\beta}\right)\,,
\end{eqnarray}
where
\vskip-.6cm
\begin{eqnarray}
\label{eq:subgrid_tensor_beta_Bardina_total}
\tau_{ij}^{\beta}\,=\,^{^{^{V}}}\!\overline{f}_i\,^{^{{S}}}\!\overline{u}_j\,-\,^{^{^{V}}}\!\widetilde{\overline{f}}_i\,^{^{{S}}}\widetilde{\overline{u}}_j\,,
\end{eqnarray}
and $\delta_j$ is the usual finite difference operator, as defined by Williams \cite{Williams_1969}. It is given by:
\vskip-.6cm
\begin{eqnarray}
\label{eq:finite_difference_operator_}
\delta_j\left(f_i\right)=\frac{1}{\Delta x_j}\left({{f_i}\,}_{i,j+1/2,k}-\, {{f_i}\,}_{i, j-1/2,k}\right).
\end{eqnarray}

In conclusion, the filtered convection-diffusion equation is obtained through the application of the Schumann box filter \cite{Schumann_1975}. While approximating the nonlinear convective term, relations between singly and doubly filtered variables arise. The scale similarity hypothesis can, then, be naturally applied. Hence, a modeled term that depends on both an eddy viscosity model (for $\tau_{ij}^\alpha$) and a scale similarity model (for $\tau_{ij}^\beta$) is achieved.

\section{MIXED MODELING: EXTENSION TO THE INCOMPRESSIBLE \break NAVIER-STOKES EQUATIONS}
\label{sec: Mixed_modeling}

The mathematical methodology described in Section \ref{sec:Sem_nome_por_enquanto} is extended to the incompressible Navier-Stokes equations. Firstly, in Section \ref{subsec:Conservation_of_mass}, the averaged conservation of mass is obtained. Secondly, in Section \ref{subsec:Conservation_of_momentum}, the equations for the conservation of filtered momentum are obtained. Finally, the subgrid-scale stress tensor is analyzed and the implemented models are defined.

\subsection{Conservation of mass}
\label{subsec:Conservation_of_mass}

The incompressibility condition $\partial / \partial x_i (u_i)=0$ is integrated over one grid cell $V$. Applying Gauss' theorem yields: 
\vskip-.6cm
\begin{eqnarray}
\label{eq:filtered_conservation_of_mass_}
\delta_j\,^{^S}\!\overline{u}_j=0\,.
\end{eqnarray}

\subsection{Conservation of momentum}
\label{subsec:Conservation_of_momentum}

The convection-diffusion equation (\ref{eq:volume_averaged_convection_diffusion_resolved_}) does not contain a pressure term. However, the contribution of the pressure can simply be added to this equation as a sink term. The averaged pressure term is given by
\vskip-.6cm
\begin{eqnarray}
\label{eq:pressure_term}
^{^{^{^{^{^V}}}}}\!\!\overline{\frac{\partial}{\partial x_i}p\,\delta_{ij}}=\frac{|S|}{|V|}\,^{^{^S}}\!\overline{p\,\delta_{ij}}=\delta_i\left(^{^{^{S}}}\!\overline{p\,\delta_{ij}}\right),
\end{eqnarray}
where $p$ and $\delta_{ij}$ are the kinematic pressure and the Kronecker delta, respectively. Here, the volume-averaged pressure term is rewritten using Gauss' divergence theorem. As before, Einstein's summation convention is used and the finite difference operator denoted by $\delta_i$ is applied to represent the fluxes through all surfaces. 
  
The pressure term is added to the convection-diffusion equation (Eq.\ (\ref{eq:volume_averaged_convection_diffusion_resolved_})) as a sink term. The physical variable $f_i$ is substituted by the velocity field $u_i$. Moreover, the diffusion coefficients $D$ and $D_e$ are substituted by the kinematic viscosities $\nu$ and $\nu_e$, respectively. The former is the fluid kinematic viscosity, while the latter is the kinematic viscosity related to the turbulence, i$.$e$.$, the eddy viscosity. Consequently, the filtered conservation of momentum for incompressible fluids is obtained:
\vskip-.6cm
\begin{eqnarray}
\label{eq:volume_averaged_coservation_of_momentum_CDE}
\frac{\partial\, ^{^{V}}\!\overline{u}_i}{\partial t}+\delta_j\left(\,^{^{{V}}}\!\widetilde{\overline{u}}_i\,^{^{{S}}}\widetilde{\overline{u}}_j\right)=-\delta_i\left(^{^{^{S}}}\!\overline{p\,\delta_{ij}}\right)+\delta_j\left(\nu\,\frac{\partial\,^{^S}\!\overline{u}_i}{\partial x_j}\,\right)-\delta_j\left(\tau_{ij}^{\alpha}+\tau_{ij}^{\beta}\right).
\end{eqnarray}

The subgrid-scale stress tensor ($\tau_{ij}^{SGS}$), which models the effect of the unresolved scales on the resolved ones, is, then, represented by a combination of the $\tau_{ij}^{\alpha}$ and $\tau_{ij}^{\beta}$ stress tensors, i$.$e$.$, by a mixed model:
\vskip-.6cm
\begin{eqnarray}
\label{eq:total_subgrid_scale_stress_tensor_}
\tau_{ij}^{SGS}=\tau_{ij}^{\alpha}+\tau_{ij}^{\beta}\,.
\end{eqnarray}

Firstly, the stress tensors $\tau_{ij}^{\alpha}$ is described (see also Eq.\ (\ref{eq:subgrid_scale_stress_tensor})):
\vskip-.6cm
\begin{eqnarray}
\label{eq:subgrid_scale_stress_tensor_velocity}
\tau_{ij}^{\alpha}\approx-\nu_e\,\left(\frac{\partial\,^{^S}\!\overline{u}_i}{\partial x_j}+\frac{\partial\,^{^S}\!\overline{u}_j}{\partial x_i}\right).
\end{eqnarray}

In the current work, the eddy viscosity $\nu_e$ is approximated according to the anisotropic minimum-dissipation model (AMD) proposed by Rozema et al.\ \cite{Rozema_2015}:
\vskip-.6cm
\begin{eqnarray}
\label{eq:minimum_dissipation_model_}
\nu_e=C\,\frac{\max\{-\left(\,^{^{^S}}\!\overline{\Delta}_k\;\;\partial \,^{^S}\!\overline{u}_i/\partial x_k\right)\left(\,^{^{^S}}\!\overline{\Delta}_k\;\;\partial\,^{^S}\!\overline{u}_j/\partial x_k\right)\,S_{ij},0\}}{\left(\partial \,^{^{^S}}\!\overline{u}_m/\partial x_l\right)\left(\partial \,^{^{^S}}\!\overline{u}_m/\partial x_l\right)},
\end{eqnarray}
where $S_{ij}$ is the rate-of-strain tensor
\vskip-.6cm
\begin{eqnarray}
\label{eq:minimum_dissipation_model_2_}
S_{ij}&=&\frac{1}{2}\left(\frac{\partial}{\partial x_j}\,^{^S}\!\overline{u}_i+\frac{\partial}{\partial x_i}\,^{^S}\!\overline{u}_j\right),
\end{eqnarray}
and $\,^{^{^S}}\!\overline{\Delta}_k$ is the filter width in the $k$-direction  of the surface filter. The constant is set to $C=0.3$ for a central second-order accurate spatial discretization method (see Rozema et al.\ \cite{Rozema_2015}). The AMD model is successfully tested, for instance, on turbulent channel flows discretized on anisotropic grids (see Rozema et al.\ \cite{Rozema_2015} and Rozema \cite{Rozema_2015_PhD}). As far as the authors are aware, the AMD model has never been tested in combination with other models.

Secondly, the stress tensor $\tau_{ij}^{\beta}$ is described (see also Eq.\ (\ref{eq:subgrid_tensor_beta_Bardina_total})):
\vskip-.6cm
\begin{eqnarray}
\label{eq:subgrid_tensor_beta_Bardina_total_velocity}
\tau_{ij}^{\beta}\,=\,^{^{{V}}}\!\overline{u}_i\,^{^{{S}}}\!\overline{u}_j\,-\,^{^{{V}}}\!\widetilde{\overline{u}}_i\,^{^{{S}}}\widetilde{\overline{u}}_j\,.
\end{eqnarray}
The tensor $\tau_{ij}^{\beta}$ can be interpreted as a scale similarity model proposed by Bardina et al.\ \cite{Bardina_1983}\break for a double decomposition of the stress tensor, where a volume and a surface average are employed (note that the Bardina model contains only volume averages: $\overline{u}_i\,\overline{u}_j-\widetilde{\overline{u}}_i\,\widetilde{\overline{u}}_j$). 

The application of a double decomposition of the stress tensor is supported by the work of Moin and Kim \cite{Moin_1982}. They state that, for second-order accurate spatial discretization schemes, an explicit calculation of the Leonard term is not justifiable since this term and the truncation error of the spatial discretization scheme are of the same order of magnitude.

In conclusion, the minimum-dissipation-Bardina mixed model is achieved in a mathematically consistent way for second-order accurate spatial discretization schemes. This model specifically combines the properties of a functional and a structural model due to their complimentary nature. Moreover, the achieved mixed model is in agreement with the $\textit{ad}$ $\textit{hoc}$ mixing approach applied by Bardina et al.\ \cite{Bardina_1983}, which sums the contributions of the Bardina model with an eddy viscosity model, such as the Smagorinsky model.

\section{CONCLUSIONS}
\label{sec: Conclusions}

The volume-balance procedure proposed by Schumann \cite{Schumann_1975} for the Navier-Stokes equations is thoroughly studied in order to achieve a mathematically consistent way of mixing the functional minimum-dissipation model \cite{Rozema_2015} and the structural scale similarity model of Bardina et al.\ \cite{Bardina_1983}. 

Firstly, the convection-diffusion equation is filtered according to the Schumann box filter \cite{Schumann_1975}. The Gauss' divergence theorem is then applied and the surface average of the\break nonlinear convective term is decomposed into a computable convective term plus a residual. The latter is approximated by the subgrid-scale stress tensor according to the eddy viscosity approach.

Secondly, the computation of the convective term requires an interpolation. This interpolation can be viewed as a filter; hence it leads to double-filtered variables. Therefore, the scale similarity hypothesis is applied. Thus, a mixed approach between an eddy viscosity model and a scale similarity model is achieved.

In this paper, the procedure is first described for a convection-diffusion equation. Thereafter, it is extended to the Navier-Stokes equations. The resulting subgrid-scale stress tensor is modeled by a mixture of the anisotropic minimum-dissipation model of Rozema et al.\ \cite{Rozema_2015} and the scale similarity model of Bardina et al.\ \cite{Bardina_1983}. In this way, a mathematically consistent mixed model is achieved.

\end{document}